\documentclass[aps,preprint,prd,showpacs,nofootinbib]{revtex4}

\usepackage{latexsym}
\usepackage{amsmath}
\usepackage{graphicx}
\usepackage{subfigure}
\usepackage{dcolumn}
\usepackage{bm}
\usepackage{amssymb}
\usepackage{color}
\usepackage{float}
\usepackage[colorlinks,linkcolor=magenta,anchorcolor=cyan,citecolor=blue]{hyperref}

\def\be{\begin{equation}}
\def\ee{\end{equation}}
\def\ba{\begin{aligned}}
\def\ea{\end{aligned}}

\def\lf{\left}
\def\rt{\right}
\def\pt{{\partial}_t}
\def\pxi{{\partial}_i}
\def\pxj{{\partial}_j}
\def\pxk{{\partial}_k}

\begin{document}

\title{ Populating the landscape in an inhomogeneous Universe }

\author{Pu-Xin Lin$^{1,2}$\footnote{linpiaoxin17@mails.ucas.ac.cn}}
\author{Yun-Song Piao$^{1,2,3,4}$\footnote{yspiao@ucas.ac.cn}}

\affiliation{$^1$ School of Physics, University of Chinese Academy
of Sciences, Beijing 100049, China}

\affiliation{$^2$ Institute of Theoretical Physics, Chinese
    Academy of Sciences, P.O. Box 2735, Beijing 100190, China}

\affiliation{$^3$ School of Fundamental Physics and Mathematical
    Sciences, Hangzhou Institute for Advanced Study, UCAS, Hangzhou
    310024, China}

\affiliation{$^4$ International Center for Theoretical Physics
    Asia-Pacific, Beijing/Hangzhou, China}

\begin{abstract}

The primordial Universe might be highly inhomogeneous. We perform
the 3+1D Numerical Relativity simulation for the evolution of
scalar field in an initial inhomogeneous expanding Universe, and
investigate how it populates the landscape with both de Sitter
(dS) and AdS vacua. The simulation results show that eventually
either the field in different region separates into different
vacua, so that the expanding dS or AdS bubbles (the bubble wall is
expanding but the spacetime inside AdS bubbles is contracting)
come into being with clear bounderies, or overall region is dS
expanding with a few smaller AdS bubbles (which collapsed into
black holes) or inhomogeneously collapsing.

\end{abstract}

\maketitle

\section{Introduction}

It has been widely thought that the inflation
\cite{Guth:1980zm,Linde:1981mu,Albrecht:1982wi,Starobinsky:1980te,Sato:1980yn,Fang:1980wi},
which may be well approximated by a de Sitter (dS) spacetime,
should happen at the early epoch of our Universe. The current
accelerated expansion of our Universe also suggests that it has a
dS-like dark energy referred as the cosmological constant.
However, a stable dS state seems not favorable in the String landscape
\cite{Bousso:2000xa,Susskind:2003kw}, which if exists, might be
extremely rare, see \cite{Ooguri:2006in,Obied:2018sgi} for the
swampland conjecture. In contrast, it is easy to construct Anti-dS
(AdS) vacua, \cite{Bousso:2000xa,Danielsson:2009ff}, see
e.g.\cite{Piao:2004me,Piao:2004hr,Garriga:2013cix,Blanco-Pillado:2019tdf,Li:2019ipk,Ye:2020btb,Ye:2020oix}
for the implications of AdS vacua on early Universe.

In such a landscape (AdS and dS vacua coexist), see Fig.\ref{potential}, whether
it is possible for our Universe to evolve to the corresponding dS
vacua and whether it is possible for it to stay in a dS state
consistent with the current observations is not obvious. Thus how
to populate the landscape, especially how our Universe started
from a dS-like inflation when AdS vacua exist, has still been a
concerned issue. It has been showed in
Refs.\cite{Coleman:1977py,Coleman:1980aw} that in an effective
potential with multiple vacua, the nucleation of bubbles with
different vacua can spontaneously occur, see also
e.g.\cite{Blau:1986cw,Lee:1987qc,Linde:1991sk,Easther:2009ft,Brown:2011ry,Braden:2018tky}.
Recent
Refs.\cite{Blanco-Pillado:2019xny,Huang:2020bzb,Wang:2019hjx} have
also reported the possibility that a large velocity fluctuation of
the scalar field pushes a region of field over the potential
barrier.

However, it is usually speculated that the primordial Universe is
highly inhomogeneous, i.e. the scalar field or spacetime metric has
large inhomogeneities before a region of space arrived at certain
vacuum. The large inhomogeneities might also be present in
multi-stream inflation \cite{Li:2009sp,Li:2009me}, in which the
inflaton field rolled along a multiple-branch path, so that the
homogeneities might hardly be preserved after bifurcations, see
also recent \cite{Cai:2021hik}. Recently, in the studies concerning large inhomogeneities, Numercial Relativity (NR), see
\cite{Lehner:2014asa,Cardoso:2014uka,Palenzuela:2020tga} for
recent reviews, has become a powerful and indispensable tool
\cite{Giblin:2015vwq,Macpherson:2016ict,Macpherson:2018btl,Giblin:2019nuv,Kou:2019bbc},
see also its application to the beginning of inflation
\cite{East:2015ggf,Clough:2016ymm,Clough:2017efm,Aurrekoetxea:2019fhr,Joana:2020rxm},
cosmological bubble collisions
\cite{Johnson:2011wt,Wainwright:2013lea,Wainwright:2014pta,Johnson:2015gma}, cosmological solitons\cite{Nazari:2020fmk} and primordial black holes \cite{deJong:2021bbo}.

It is interesting and significant to perform the 3+1D NR
simulation in an initial inhomogeneous Universe to investigate how
the scalar field populates the landscape. We will work with a
highly inhomogeneous Universe that is initially expanding and a scalar field (its
effective potential has both dS and AdS vacua), and numerically
evolve it with modified NR package
GRChombo\footnote{http://www.grchombo.org\\https://github.com/GRChombo}
\cite{Clough:2015sqa}. This paper is outlined as follows. In
Section II, we present the model and initial conditions. In
Sections III, we present the simulation results and discuss the
relevant implications. We conclude in Section V. We will set
$c=8\pi G =1$. Throughout the paper, we will set the reduced Planck mass $\tilde{M}_{Pl}=1$.

\section{The model and initial conditions}

\begin{figure}[H]
    \centering
    \includegraphics[width=8cm]{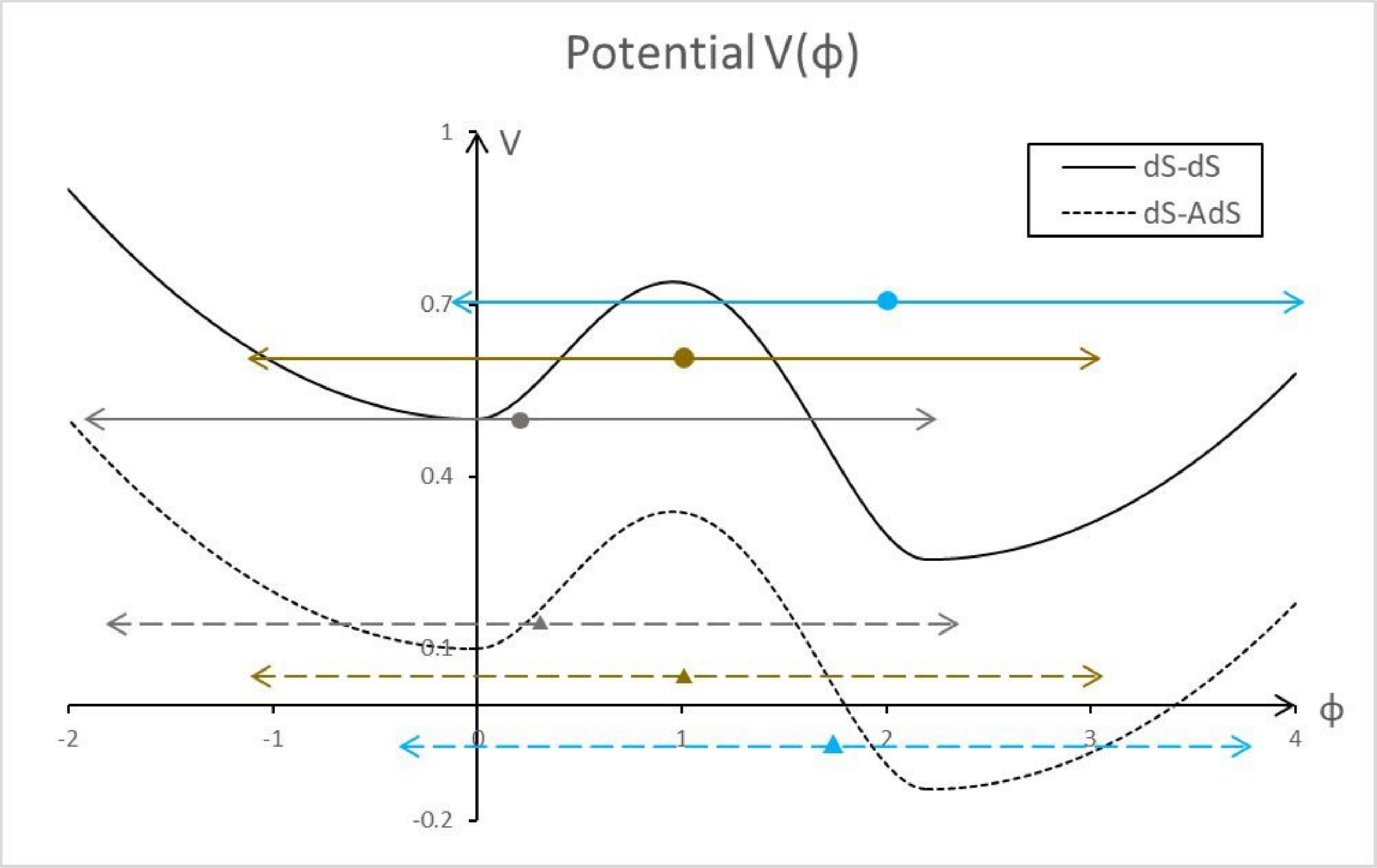}
\caption{The upper panel is the dSdS potential (both minima are
dS-like), the lower panel is the dSAdS potential (one minimum is
dS-like, the other is AdS-like). Lengths of the arrow lines
represent the initial amplitude of $\phi$. The brown, grey, blue
colored lines and dotted lines correspond to different sets in our
simulations, i.e.the 1st, 2nd and 3rd set, respectively. }
    \label{potential}
\end{figure}

In an effective theory, the string landscape might correspond to a
complex and rugged potential. However, for simplicity, we set the
potential $\sim \phi^2$ around its minima \footnote{This helps to
ease the computational cost of relaxing the initial condition.},
which are separated by a fourth-order polynomial barrier, \be
V\lf(\phi \rt) = \left\{\begin{aligned}
    &\frac{1}{2}m_1^2 { \lf( \phi - \phi_1 \rt) }^2 +V_1, &\phi<\phi_1,\\
   &\lambda\lf(\phi^2+{1\over 2\lambda}\rt)^2-\phi^3-{1\over 4\lambda}+V_1, \quad &\quad \phi_1\leq\phi\leq\phi_2, \\
    &\frac{1}{2}m_2^2 { \lf( \phi - \phi_2 \rt) }^2+V_2,&\phi_2<\phi,
\end{aligned}\right.
\label{V}\ee see Fig.\ref{potential}. The minima of the
fourth-order polynomial are at $\phi_1=0$ and
$\phi_2=\frac{3+\sqrt{9-32\lambda}}{8 \lambda}$ $(\lambda <
\frac{9}{32})$, respectively, at which the potential is
differentiable. In the simulation, we will fix $m_1^2=m_2^2=0.2$
and $\lambda=0.2375$.

The initial inhomogeneity of the scalar field $\phi$ is regarded as  \be \lf.\phi
\rt|_{t=0} = \phi_0 + \Delta \phi \sum_{\vec{x}=x,y,z}\cos \lf(
\frac{2\pi \vec{x}}{L} \rt),\ee  similar to that in
Refs.\cite{East:2015ggf,Clough:2016ymm}, where $\vec{x}$ is the
spatial coordinate, $\Delta \phi$ is the amplitude of initial
inhomogeneity, while the length of the simulated cubic region is
$L=4$. The initial
expansion rates for the dSdS and dSAdS scenarios are $H_{init} =
0.7, 0.6$ respectively, corresponding to Hubble radii $H^{-1} =
1.43, 1.67 < L$, i.e. the initial scale of inhomogeneity is
superhorizon.

In light of the potential in (\ref{V}), we classify the scenarios
simulated as dSdS (both vacua are dS-like) and dSAdS (one is
dS-like and the other is AdS-like). We will consider
simulations of 3 cases (for dSdS and dSAdS,
respectively)\footnote{The result is labeled by the scenario it
belongs to (dSdS or dSAdS) followed by the different case numbers,
e.g. dSdS-1.}, with $\Delta\phi = 0.7$ but different average field
value $\phi_0=0.2,1.0,1.7$ for dSdS ($\phi_0=0.3,1.0,2.0$ for dSAdS), where $\phi_0=0.2,1.7 (0.3, 2.0)$ indicates that
the initial distribution of $\phi$ is biased towards one of the
vacua, see Fig.\ref{potential}. Here, the inhomogeneity considered clearly exceeds
the perturbative level. However, during the very early stage of
the Universe, the initial inhomogeneity might arise from large quantum
fluctuations with $\Delta\phi\simeq H$, where $H\sim
\mathcal{O}(1)$. In addition, the String landscape conjectures
bounds on the scalar field excursion, e.g., in \cite{Agrawal:2018own, Ooguri:2006in} $|\Delta \phi| < \mathcal{O}(1)$, which is also consistent with our model.

Appendix A shows a brief review on NR based on BSSN \cite{Baumgarte:1998te,Shibata:1995we} and the symbols and conventions in our paper.
We set the initial values of BSSN parameters as
$\widetilde{\gamma}_{ij}=\delta_{ij}$, $\widetilde{A}_{ij}=0$,
and the initial spatial expansion uniform
($K=const. <0$) and $\dot{\phi}=0$, which naturally satisfy the momentum constraints. The
Hamiltonian constraint is then solved by relaxing $\chi$ from the initial value $\chi=1$ with the parabolic equation ${\partial}_t \chi =\mathcal{H}$. This equation
is iterated until it converges (suggesting $\pt \chi = \mathcal{H}
=0$). The resolution of the simulation is $32 \times 32 \times 32$ along the x,y,z axes on the coarsest level with up to 3 levels of AMR regridding.

\section{Results and analysis}

\begin{figure}[!hbt]
    \centering
    \includegraphics[width=16cm]{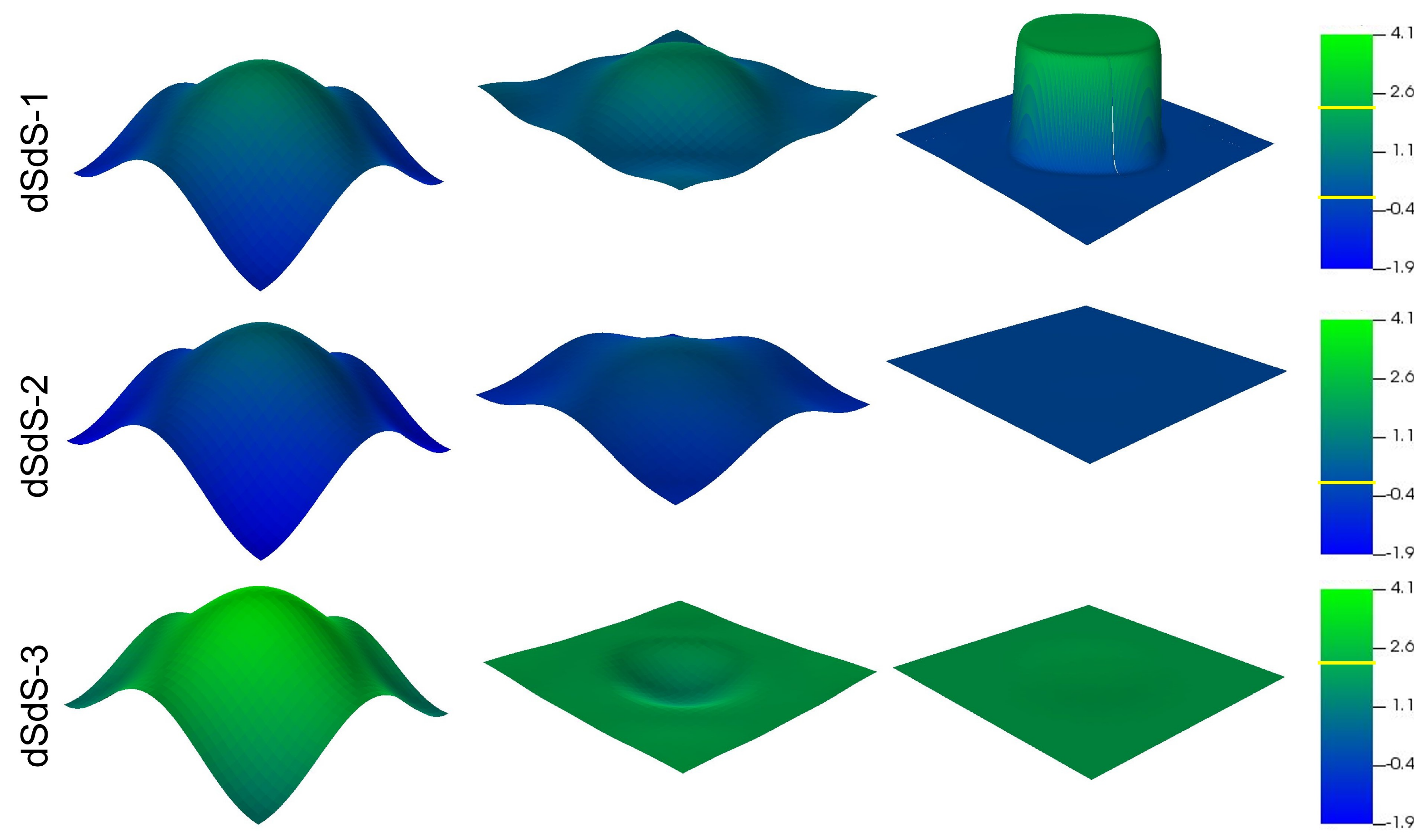}
    \caption{Results of $\phi$ at $x-y$ slices at different timesteps
        (proceeding from left to right) for dSdS-1,2,3. The colorbar and
        the elevation height show the value of $\phi$. The yellow line on
        the colorbar marks the converged values of $\phi$. }
    \label{dSdS123_phi}
\end{figure}

\begin{figure}[!hbt]
    \centering
    \includegraphics[width=16cm]{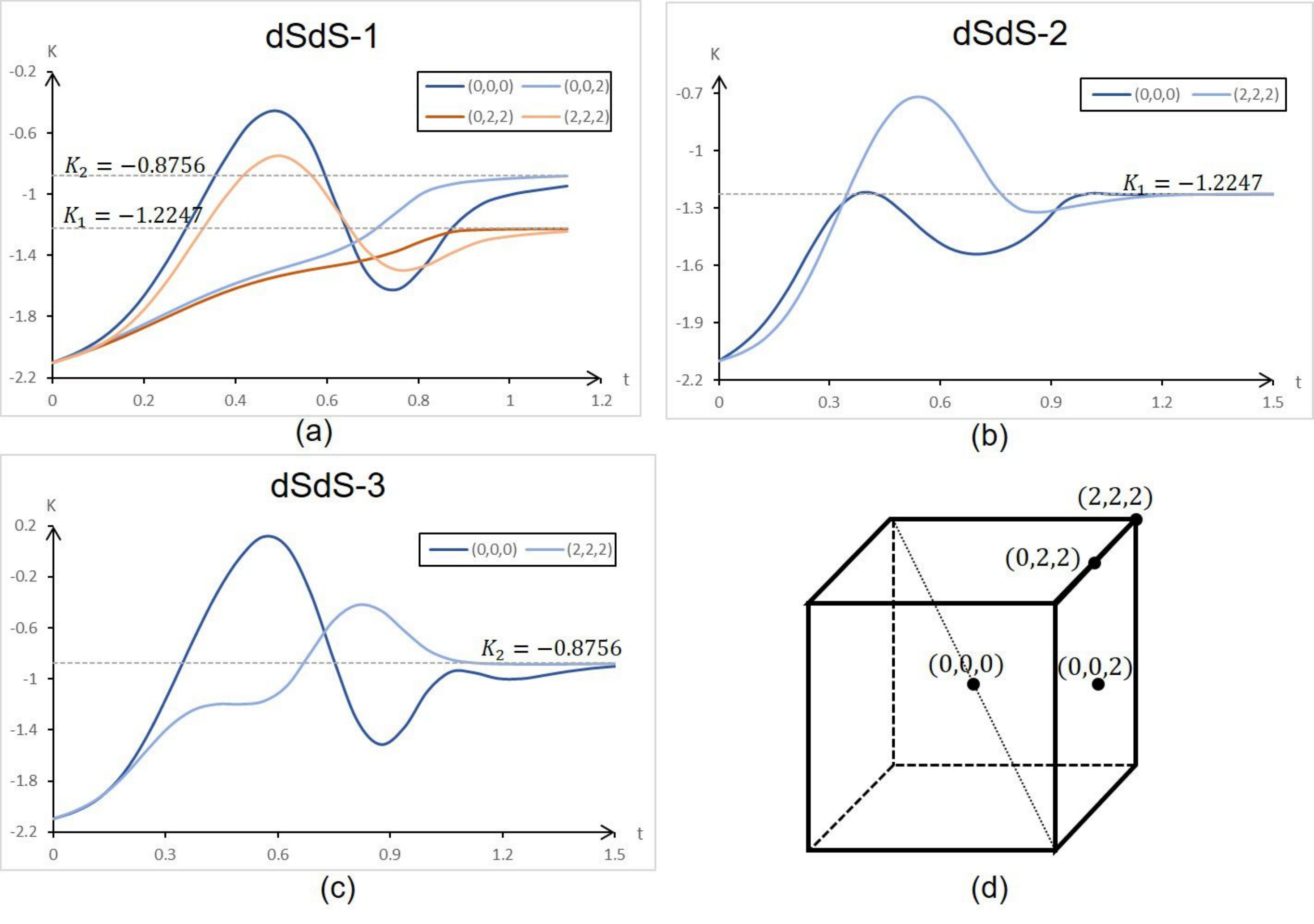}
    \caption{The evolution of $K$ for dSdS-1,2,3. $K_1$ and $K_2$
        correspond to the local Hubble rate at $\phi_1$ and $\phi_2$,
        respectively (noting both $V(\phi_1),V(\phi_2)>0$ for dSdS in
        Fig.\ref{potential}). (d) shows the labelling of the positions in (a)(b)(c). These positions are at the bulk (center) of the vacua or where
        these regions intersect the simulation boundary, and best capture
        the physics of corresponding vacua.}
    \label{dSdS_K}
\end{figure}

\begin{figure}[!hbt]
    \centering
    \includegraphics[width=16cm]{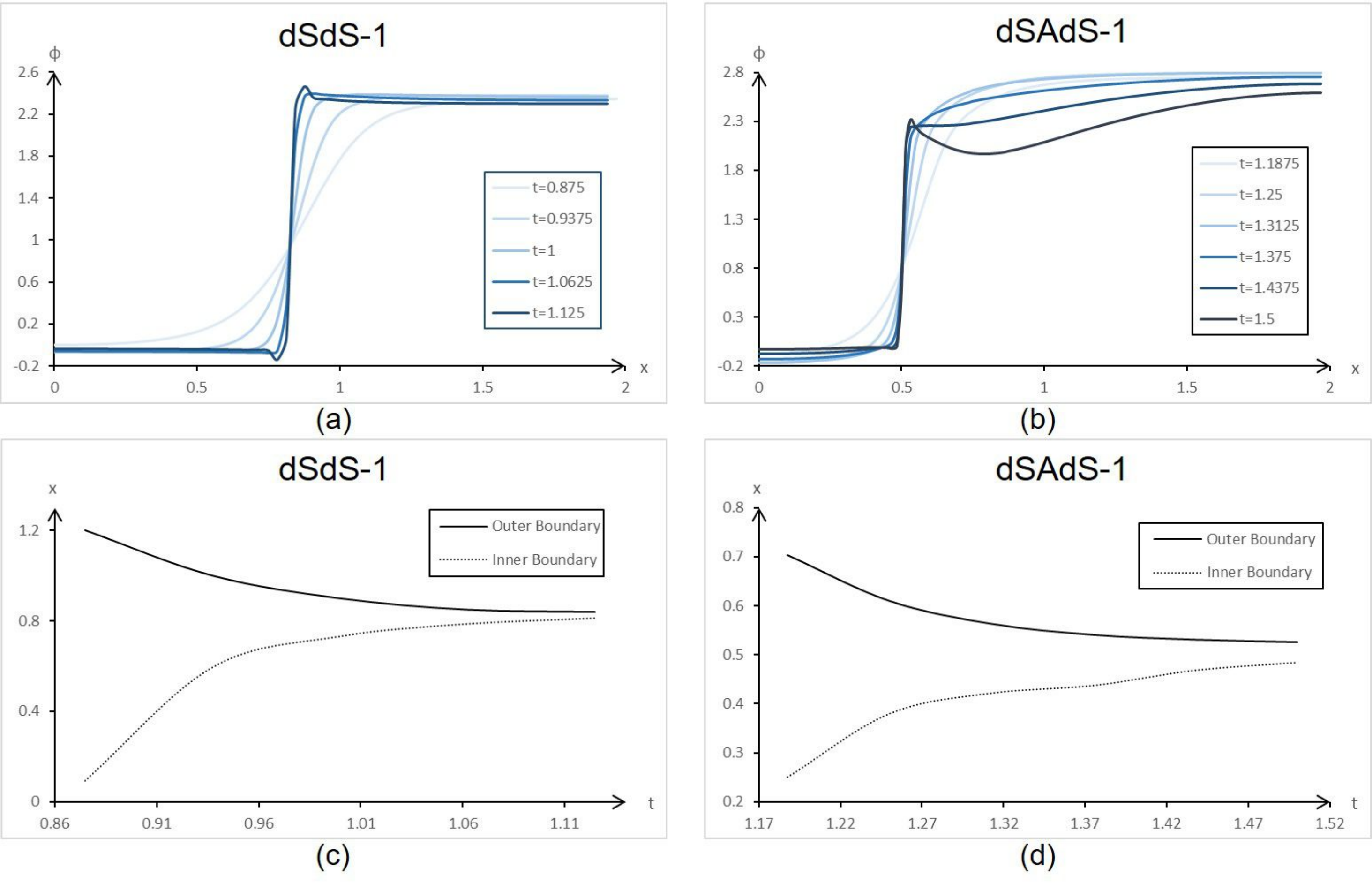}
\caption{Upper row: the value of $\phi$ along the
$x$-axis, showing the different vacua regions with $\phi =
2.2$ and $\phi = 0$ separated by a bubble wall. The darker lines
show the field configurations for later time. Second row:
positions of the inner and outer boundaries of the bubble wall. It
is clear that the width of the bubble wall in comoving coordinates
shrinks with time, which indicates that the position of the
bubble wall freezes.}
    \label{Bubble Wall}
\end{figure}

\begin{figure}[!hbt]
    \centering
    \includegraphics[width=16cm]{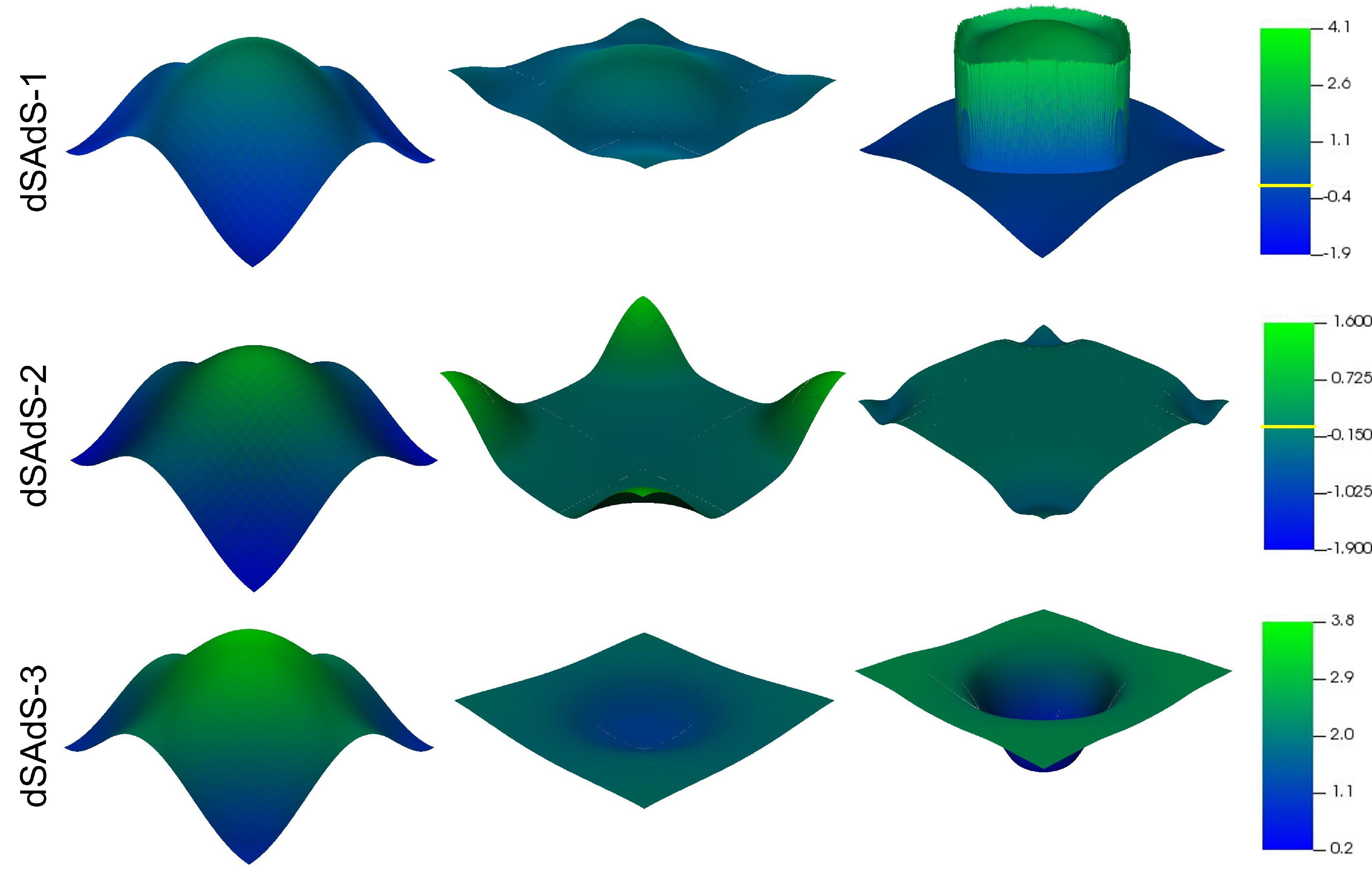}
    \caption{Results of $\phi$ at $x-y$ slices at different timesteps
        (proceeding from left to right) for dSAdS-1,2,3. The colorbar and
        the elevation height show the value of $\phi$. The yellow line on
        the colorbar marks the converged values of $\phi$.}
    \label{dSAdS123_phi}
\end{figure}

\begin{figure}[!hbt]
    \centering
    \includegraphics[width=16cm]{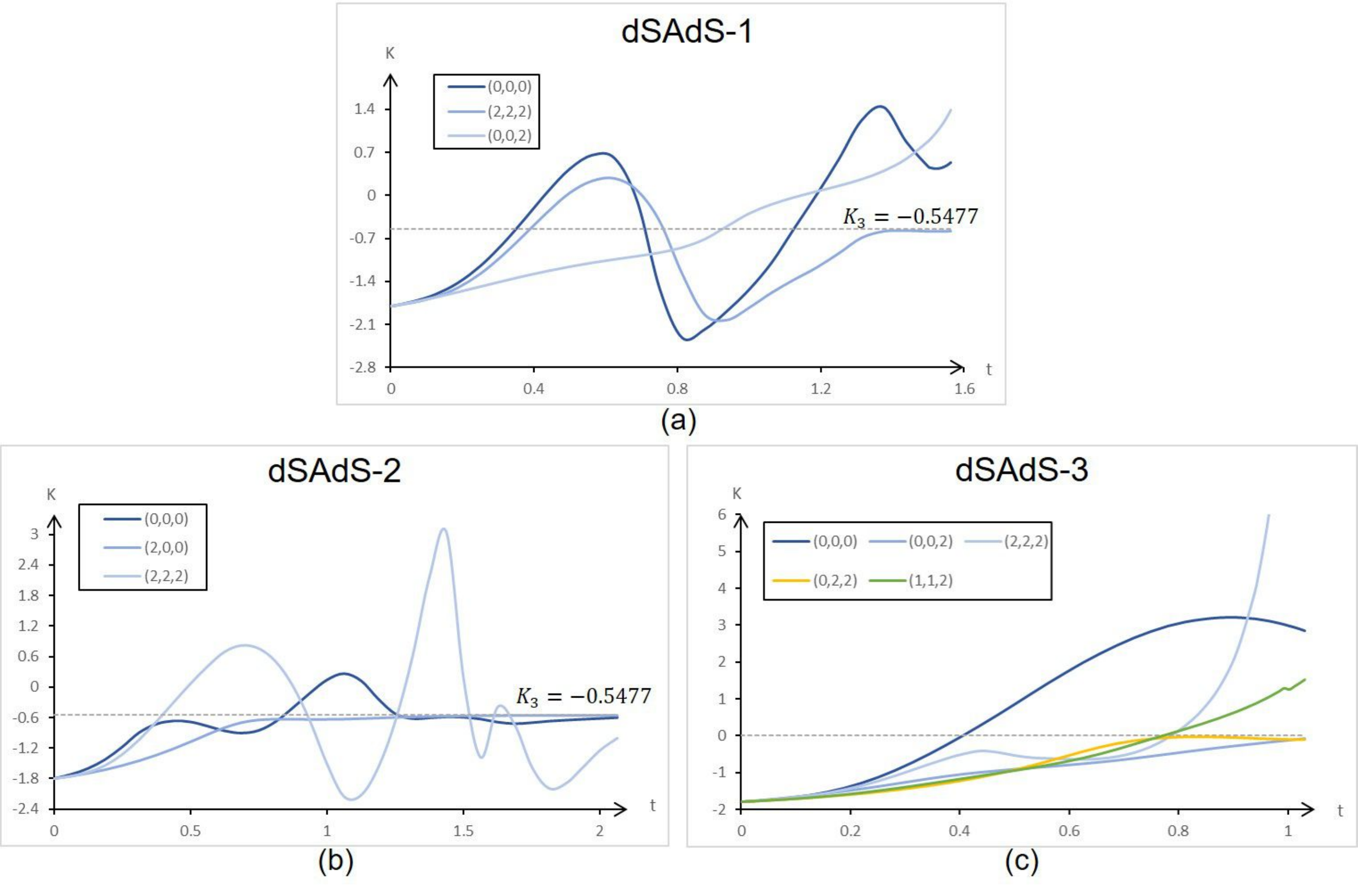}
    \caption{The evolution of $K$ for dSAdS-1,2,3. $K_3$ corresponds
        to the local Hubble rate at $\phi_1$ (noting $V(\phi_1)>0$ and
        $V(\phi_2)<0$ for dSAdS in Fig.\ref{potential}). The positions are labelled in accordance with Fig.\ref{dSdS_K} (d). }
    \label{dSAdS_K}
\end{figure}

We will perform the NR simulations with modified GRChombo package
to investigate how the field populates the landscape in
Fig.\ref{potential} in an inhomogeneous Universe that is initially expanding.

As a contrast, we first consider a landscape consisting of only dS
vacua. We show the evolutions of $\phi$ at certain regions for
dSdS-1,2,3 in Fig.\ref{dSdS123_phi}. The field initially underwent
a rapidly oscillating phase. However, eventually, for dSdS-1 the
field in different spatial region will separate into different
vacua, and the dS bubbles come into being with clear boundries,
while for dSdS-2,3, the overall region will be in a nearly homogeneous
dS expansion.

The Hubble rate at a local homogeneous region is
 \be
H_{local}=\lim\limits_{V\rightarrow 0}\sqrt{{1\over
V}\int{\rho(\phi)\over 3} dV}=-{K\over 3},
\label{local_H}
\ee
In Fig.\ref{dSdS_K}, for dSdS-1, the local Hubble rate will be
$H_{local}=({V(\phi_{1,2})\over 3})^{1/2}$, i.e. that at different
vacua $\phi_1$ and $\phi_2$, respectively, and for dSdS-2,3,
$H_{local}$ will be identical eventually at all region, i.e. a
homogeneous dS expansion. The result is consistent with
Fig.\ref{dSdS123_phi}.

In our simulation results for dSdS-1, eventually the dS bubbles
will emerge in high-energy dS background. It is well-known that if
the radius of the bubble is larger than the Hubble radius of the
background, $r\geqslant H^{-1}$, the bubble wall will expand with
the background. In Fig.\ref{dSdS123_phi}, we see that the position
of the bubble wall is frozen, suggesting that the dS bubble is in fact
expanding with the background. However, the condition $r>1/H$ is
not strictly satisfied in our simulation, the Hubble length
$H^{-1}$ of background in dSdS-1 is 2.45, which is comparable but
not less than the the radius $r\simeq 1$ of bubble.

It is more interesting to investigate the dSAdS landscape in
Fig.\ref{potential}. We show the evolutions of $\phi$ at
different regions for dSAdS-1,2,3 in Fig.\ref{dSAdS123_phi}. Results show that dSAdS behaved similarly to dSdS only at the initial
stage of the evolution.

\begin{figure}[H]
    \centering
    \includegraphics[width=16cm]{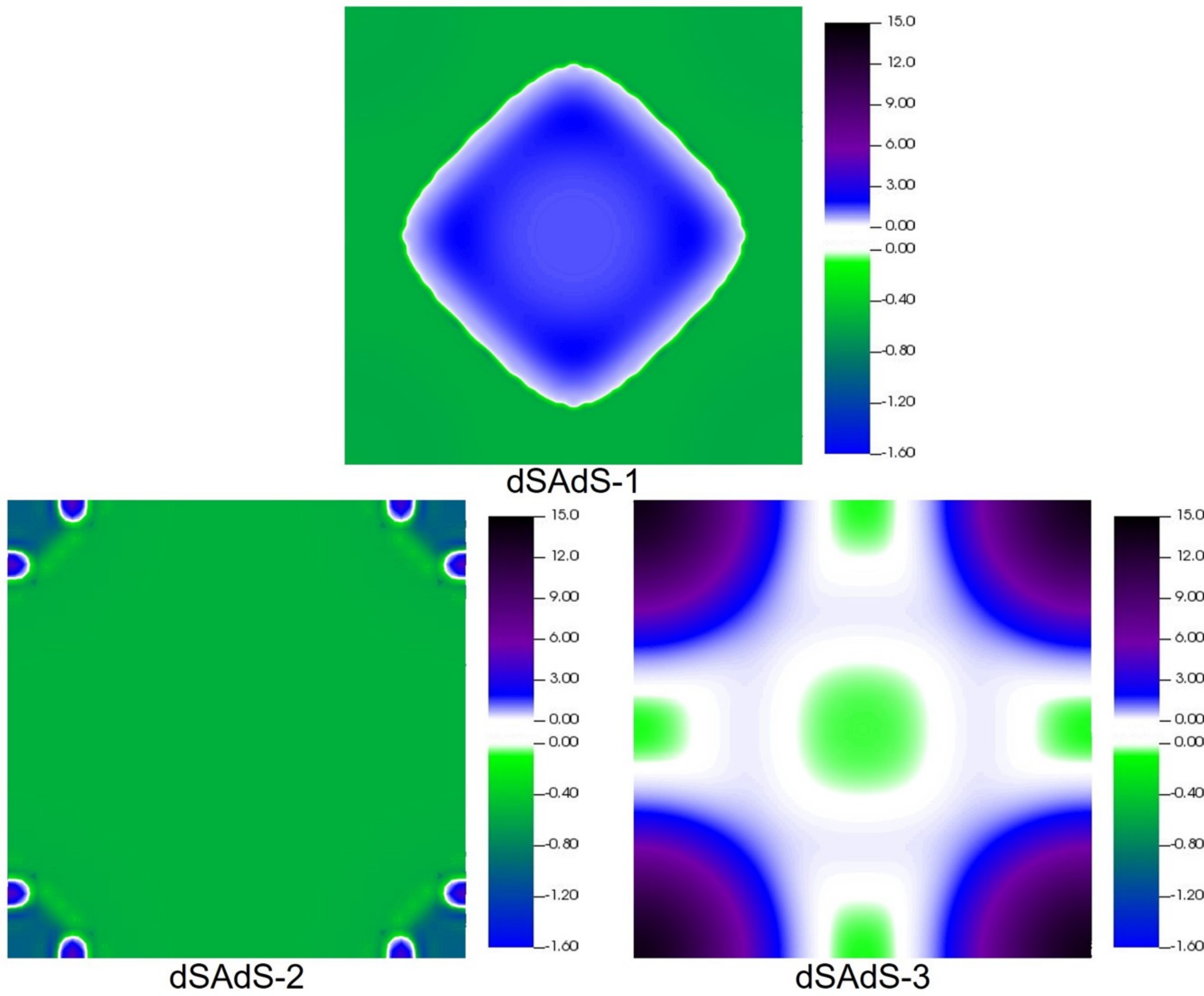}
\caption{The values of $K$ for dSAdS-1,2,3 at final timesteps.
There are two sets of colorbars: the upper colorbar shows the
value of $K$ for $K>0$ (the contracting region), while the lower
colorbar shows that for $K<0$ (the expanding region), the
two regions are seperated by walls where $K=0$ colored in white.}
    \label{dSAdS123_K}
\end{figure}

It is significant to check the local expansion rate. In
Fig.\ref{dSAdS_K}\footnote{In our simulation, the calculations
will stop whenever a single point in space diverged.}, for
dSAdS-1, some regions eventually converged to $H_{local}=const.>0$
(the dS expansion), while other regions have $H_{local}<0$. In Fig.\ref{dSAdS123_K}, these contracting AdS regions will always collapse. Ref.\cite{Felder:2002jk}
investigated a homogeneous case with $V_{min} < 0$ and showed the
diverging property of $H$ once it crosses to the negative side,
which indicates the final fate of the AdS bubbles in our
simulation. For dSAdS-2, after the
initial oscillation, the overall region will have a nearly
homogeneous dS expansion, except for a few smaller AdS bubbles,
see Fig.\ref{dSAdS123_K}. Thus our results show that the expanding
dS regions may be present eventually, even if the AdS vacua exist.
For dSAdS-3, the story is different. Due to the rapid collapse of the
AdS spacetime, the numerical code is unable to evolve the system
after the collapsing regions run into ``singularities" somewhere. In Fig.\ref{dSAdS123_K}, we see that some regions with $K<0$ are left
at the end of the simulation. However, these regions cannot
evolve to a stable dS spacetime, because the profile of $\phi$ has
crossed the potential barrier and fallen in the range of the AdS
minima, see the 3rd row of Fig.\ref{dSAdS123_phi}. The corresponding AdS
vacua will eventually stop the expansion of these regions and
convert them into collapsing spacetime. We thus conclude that
the overall region will eventually be AdS-like, resulting in an
inhomogeneous collapse.

In our simulation result for dSAdS-1, see Fig.\ref{Bubble Wall},
eventually the position of the bubble wall is frozen, suggesting
that the wall of AdS bubble is expanding with bakground, so such
AdS bubbles correspond to the separated Universes, but the
spacetime inside AdS bubbles is contracting (confirming the
argument of Ref.\cite{Abbott:1985kr}). The radius of AdS bubbles
is approximately $r< 1/H\simeq 5.48$. Again, as in dSdS-1, the condition
$r>1/H$ is not satisfied. The contracting AdS bubble might be
relevant to our Universe
\cite{Piao:2004me,Piao:2009ku,Johnson:2011aa,Garriga:2013cix}, if
a nonsingular bounce happened, which might explain the large-scale
CMB power deficit \cite{Piao:2003zm,Liu:2013kea,Cai:2017pga}.
While in dSAdS-2, the AdS bubbles have its radius $r\ll 1/H$,
which (to observers outside the bubbles) will then collapse into
black holes\footnote{It has been argued in
Ref.\cite{Clough:2016ymm} that the large inhomogeneities of the
scalar field may create black holes. However, our case is
different, the black holes result from the collapse of AdS
bubbles.}, see also \cite{Blanco-Pillado:2019tdf, Garriga:2015fdk}.

\section{Conclusions}

It is usually speculated that the primordial Universe is highly
inhomogeneous, i.e. the scalar field or spacetime metric has large
inhomogeneities. How the landscape is populated in an inhomogeneous
Universe is still a significant question.

In an inhomogeneous Universe that is initially expanding, we perform the
3+1D Numerical Relativity simulations for the evolution of a scalar
field in the simplified landscape in Fig.\ref{potential}, and
investigated how the field populates such a landscape. The
simulation results showed that eventually, either
the overall region is in a nearly homogeneous dS expansion
(for dSAdS, however, a few smaller regions corresponding to AdS bubbles collapsed into black holes), or the whole region is
inhomogeneously collapsing, or the field in different spatial region
separates into different vacua, so that expanding dS and AdS
bubbles (the bubble wall is expanding but the spacetime inside AdS
bubbles is contracting) come into being with clear bounderies.

It is noted that the initial high inhomogeneities seem to amplify
the probability that different regions of Universe arrive at
different vacua. Also, the bubble wall expand with the background seems to not require that the bubble radius must be
strictly larger than the Hubble radius. Though we perform the
simulation in a simplified landscape, our results have captured
relevant physics.

The theory of inflation as a paradigm of
early universe indicates an existence of a dS or quasi-dS phase
of the Universe. Studies on the String landscape had attempted to
answer the questions of whether a dS vacua is permitted in String
theory and, if so, how can dS vacua be constructed. On the
other hand, the initial conditions of the Universe is not
necessarily (in fact, unlikely) homogeneous. Here, we
discuss the questions of how in a highly inhomogeneous
Universe (where the dS vacua is not yet populated) a patch of
spacetime evolves into the dS vacua, if they exist. We showed that the ``islands" of dS spacetime can naturally emerge
depending on the initial conditions of the field configuration.
Thus we actually suggested the possibility for the
appearance and existence of local dS spacetime that corresponds
to our Universe.

An interesting issue that follows is what
signals would we ``see" if our Universe indeed went through such
an inhomogeneous evolution before or during inflation?\\

\textbf{Acknowledgment} PXL would like to thank Gen Ye, Hao-Hao
Li, Hao-Yang Liu for helpful discussions. We also acknowledge the
use of the package GRChombo and VisIt. This work is supported by
NSFC, Nos.12075246, 11690021, and also UCAS Undergraduate
Innovative Practice Project.

\appendix

\section{A brief review on NR and BSSN formalism}

In the context of 3+1 decomposition of NR, the metric is \be
g_{00} = -{\alpha}^2+{\beta}_i {\beta}^i ,\qquad g_{0i} =
{\beta}_i ,\qquad g_{ij} = {\gamma}_{ij}, \ee where $\alpha$ is
the lapse parameter, ${\beta}^i$ the shift vector and
${\gamma}_{ij}$ the spatial metric. In order to formulate the
evolution of spacetime and the ``matter" inside as a well-posed
Cauchy problem, the system of partial differential equations should
be explicitly written in a hyperbolic form. According to BSSN
\cite{Baumgarte:1998te,Shibata:1995we}, the evolution equation are
\be {\partial}_t \chi=\frac{2}{3}\alpha \chi K-\frac{2}{3}\chi
{\partial}_k {\beta}^k+{\beta}^k {\partial}_t \chi, \ee \be
{\partial}_t \widetilde{\gamma}_{ij}=-2\alpha \widetilde{A}_{ij}+
\widetilde{\gamma}_{ik} \pxj {\beta}^k+ \widetilde{\gamma}_{jk}
\pxi {\beta}^k-\frac{2}{3} \widetilde{\gamma}_{ij} \pxk
{\beta}^k+{\beta}^k \pxk \widetilde{\gamma}_{ij}, \ee \be \pt
K=-\widetilde{\gamma}^{ij}D_i D_j \alpha +\alpha
\lf(\widetilde{A}_{ij} \widetilde{A}^{ij} + \frac{1}{3}K^2
\rt)+{\beta}^i \pxi K + 4\pi G\alpha \lf(\rho + S \rt), \ee \be
\ba
\pt \widetilde{A}_{ij} &= \chi {\lf[ -D_i D_j \alpha + \alpha \lf( R_{ij} -8 \pi \alpha S_{ij} \rt) \rt] }^{TF}+\alpha \lf( K \widetilde{A}_{ij} -2 \widetilde{A}_{il} \widetilde{A} \rt)\\
&+\widetilde{A}_{ik} \pxj {\beta}^k +\widetilde{A}_{jk} \pxi
{\beta}^k - \frac{2}{3} \widetilde{A}_{ij} \pxk {\beta}^k
+{\beta}^k \pxk \widetilde{A}_{ij}, \ea \ee \be \ba
\pt \widetilde{\Gamma}^i &= -2\widetilde{A}^{ij} \pxj \alpha +2\alpha \lf( \widetilde{\Gamma}^i_{jk} \widetilde{A}^{jk} -\frac{2}{3} \widetilde{\gamma}^{ij} \pxj K -\frac{3}{2 \chi} \widetilde{A}^{ij} \pxj \chi \rt)+{\beta}^k \pxk \widetilde{\Gamma}^i + \widetilde{\gamma}^{jk} \pxj \pxk {\beta}^i \\
&+\frac{1}{3} \widetilde{\gamma}^{ij} \pxj \pxk {\beta}^k
+\frac{2}{3} \widetilde{\Gamma}^i \pxk {\beta}^k -
\widetilde{\Gamma}^k \pxk {\beta}^i -16\pi G \alpha
\widetilde{\gamma}^{ij} S_j, \ea \ee where the tilde represent the
conformal quantities $\widetilde{\gamma}_{ij}=\chi \gamma_{ij}$,
$\widetilde{\Gamma}^i\equiv \widetilde{\gamma}^{jk}
\widetilde{\Gamma}^i_{jk}$ and $K$ is the extrinsic curvature. The
Hamiltonian and momentum constraints are \be
\mathcal{H}=\widetilde{D}^2 \chi -\frac{5}{4
\chi}\widetilde{\gamma}^{ij} \widetilde{D}_i \chi \widetilde{D}_j
\chi +\frac{\chi
\widetilde{R}}{2}+\frac{K^2}{3}-\frac{1}{2}\widetilde{A}^{ij}\widetilde{A}_{ij}-8\pi
G \rho=0, \label{Hamilton}\ee \be \mathcal{M}^i=\widetilde{D}_j
\widetilde{A}^{ij}-\frac{3}{2 \chi}
\widetilde{A}^{ij}\widetilde{D}_j \chi
-\frac{2}{3}\widetilde{\gamma}^{ij}\widetilde{D}_j K - 8\pi G \Pi
\widetilde{\gamma}^{ij} \pxj \phi=0. \label{momentum}\ee

The Klein-Gordon equation of canonical scalar field $\phi$ is
$\Box \phi = -V'$. According to BSSN, it is rewritten as
\cite{Clough:2015sqa} (with the momentum conjugate $\Pi =
\frac{1}{\alpha} \lf( \pt \phi - \beta^i \pxi \phi \rt)$)
\be \pt \Pi =\alpha
\lf( K \Pi - \Gamma^k \pxk \phi - V'\rt) + \alpha \pxi
\partial^i \phi + \pxi \phi \partial^i \alpha + \beta^i \pxi \Pi.
\ee

The gauge conditions in our simulations are the 1+log slicing and
the Eulerian gauge, \be \pt \alpha = -2 \alpha K + \beta^i \pxi
\alpha, \quad \beta^i=0,  \label{comoving_gauge} \ee , see
e.g.\cite{alcubierre2008introduction, baumgarte2010numerical,
Gourgoulhon:2007ue, Lehner:2001wq} for details.

\section{On Hamiltonian constraint and convergence test}

\begin{figure}[H]
    \centering
    \includegraphics[width=16cm]{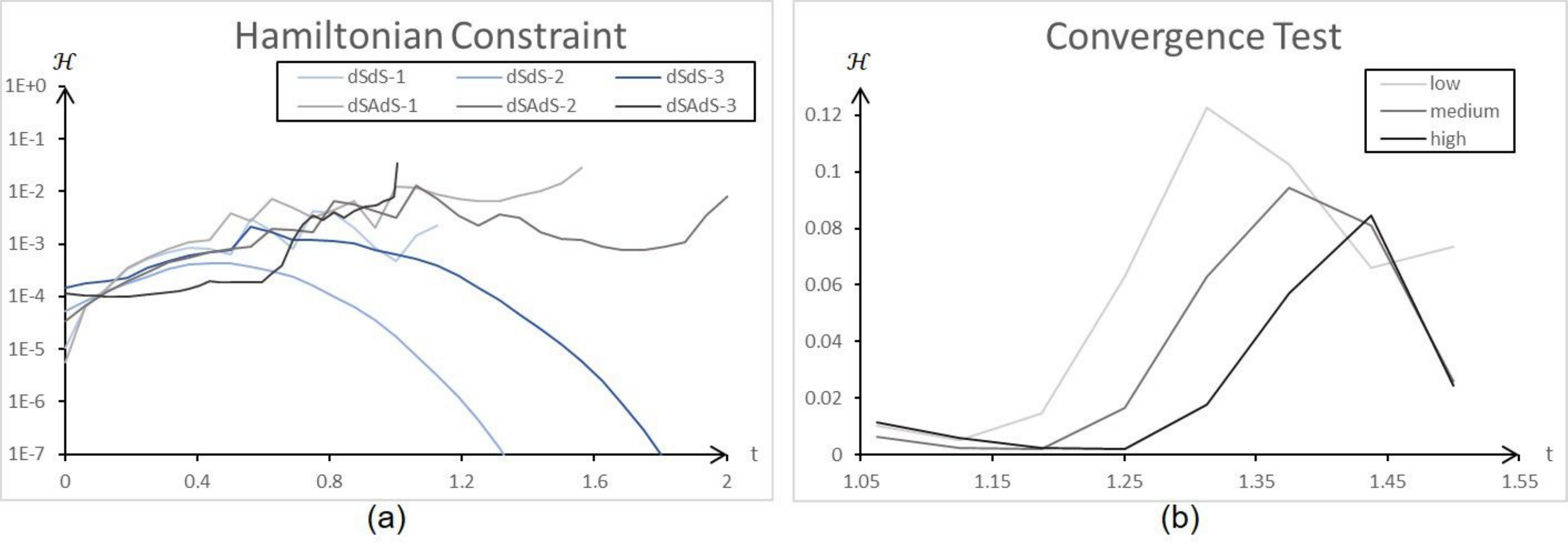}
\caption{(a): The box-averaged $L_2$-norm of the Hamiltonian
constraint $\mathcal{H}$. Due to the fact that the Hamiltonian of
different points can cross $H=0$, a fraction Hamiltonian violation
cannot be globally defined. Thus, with the data yielding the
average $H\sim O\lf(1\rt)$, we instead keep the absolute
constraint violation under $10^{-2}$. (b): A
convergence test with different grid resolutions. The presented
results shows the Hamiltonian violation of case dSAdS-1 
when the bubble wall formed. Higher resolution yields higher
pricision at the start of the bubble wall formation, afterwards,
the results under meduim and high resolutions start to converge.}
    \label{Hamiltonian_Constraint}
\end{figure}

\bibliography{references}

\end{document}